# A Recommendation Model Utilizing Separation Embedding and Self-Attention for Feature Mining

Wenyi Liu
Independent Researcher
Nanjing, China

Rui Wang
Carnegie Mellon University
Pittsburgh, USA

Yuanshuai Luo
Southwest Jiaotong University
Chengdu，China

Jianjun Wei
Washington University in St. Louis
St Louis，USA

Zihao Zhao
Stevens Institute of Technology
Hoboken, USA

Junming Huang*
Carnegie Mellon University
Pittsburgh, USA

*Abstract*—With the explosive growth of Internet data, users are facing the problem of information overload, which makes it a challenge to efficiently obtain the required resources. Recommendation systems have emerged in this context. By filtering massive amounts of information, they provide users with content that meets their needs, playing a key role in scenarios such as advertising recommendation and product recommendation. However, traditional click-through rate prediction and TOP-K recommendation mechanisms are gradually unable to meet the recommendations needs in modern life scenarios due to high computational complexity, large memory consumption, long feature selection time, and insufficient feature interaction. This paper proposes a recommendations system model based on a separation embedding cross-network. The model uses an embedding neural network layer to transform sparse feature vectors into dense embedding vectors, and can independently perform feature cross operations on different dimensions, thereby improving the accuracy and depth of feature mining. Experimental results show that the model shows stronger adaptability and higher prediction accuracy in processing complex data sets, effectively solving the problems existing in existing models.

*Keywords-Separation embedding cross network; Efficient recommendation algorithm; Self-attention mechanism*

## I. INTRODUCTION

In the context of the rapid development of science and technology today, the surge in the amount of Internet data has led to amounts of information overload, which has brought challenges to users in efficiently obtaining the required resources [1]. With the advancement of science and technology, the way of obtaining information has also undergone tremendous changes from traditional letters and newspapers to the modern Internet [2]. More and more individuals tend to obtain information through the Internet platform, and the emergence of recommendation systems plays a key role in this. It can effectively filter massive information and provide users with content that matches their needs [3]. Traditional recommendation systems, while valuable, face numerous challenges in the modern context where data is complex, dynamic, and vast in scale. Specific issues such as high-latency responses, poor scalability, and limited adaptability to real-time data updates render these traditional systems less effective in rapidly evolving domains like e-commerce and video streaming platforms. Additionally, traditional click-through rate prediction and TOP-K recommendation mechanisms often suffer from high computational complexity, large memory consumption, long feature selection time, and insufficient feature interaction, further limiting their applicability in real-world scenarios [4].

On this basis, this study proposes a recommendation system model based on a separation embedding cross-network, which aims to solve the problem that features cross technology in existing deep learning recommendation models fail to fully mine embedded vector information and is insufficient in accuracy [5]. The proposed model utilizes an embedding neural network layer to transform high-dimensional, sparse feature vectors into low-dimensional, dense embedding vectors. This transformation allows the model to efficiently encode complex feature relationships by reducing the dimensionality of input data while preserving essential information. Specifically, the embedding neural network layer assigns trainable weights to each feature, mapping them into a continuous vector space, where the previously sparse features can now interact more robustly. This transformation facilitates deeper and more precise feature interactions in subsequent layers, particularly during the independent feature cross operations across different dimensions [6]. Finally, the hidden layer matrices are summed and pooled, and the final recommendation result is obtained through the prediction layer [7]. The experimental results show that compared with the six popular typical click-through rate prediction models and five recommendation schemes based on graph neural networks, the performance of this model on three public datasets is better, specifically in terms of higher AUC values, accuracy, and recall [8].

In addition, this study also introduces a recommendation model design based on the self-attention mechanism [9]. The self-attention mechanism allows the model to focus on different parts of the input when processing sequence data, which has been used in various fields such as computer vision [10-12], thereby capturing more complex dependencies. By applying the self-attention mechanism to the recommendation system, this study aims to enhance the model's ability to understand the user's historical behavior patterns in order to provide more personalized recommendation services. The experimental results show that this model not only improves the accuracy of recommendations, but also effectively reduces training time and resource consumption.

Finally, this study also explores how to optimize the feature screening process and proposes a set of efficient feature engineering solutions. By improving the feature selection algorithm and combining domain knowledge with machine learning techniques, we designed a method that can quickly identify the most influential feature combinations. This method not only improves the speed of feature selection, but also further enhances the generalization ability and predictive performance of the model. Taken together, these improvements have jointly promoted the performance of recommendation systems in practical applications and laid a solid foundation for future personalized recommendation technologies.

## II. METHOD

At the beginning of the experimental design, we first preprocessed the data. Preprocessing includes data cleaning, standardization, normalization and feature engineering. Data cleaning is mainly to remove missing values or outliers to ensure the quality of the data. Standardization and normalization are to make the data on the same scale to avoid unstable model training due to too large or too small a value range [13]. Feature engineering refers to transforming, combining or deriving new features from the original data to enhance the expressiveness of the model [14]. For example, new features can be generated by counting the user's historical behavior, such as user activity, preference type, etc. This series of operations can be summarized as:

$$X' = \text{Preprocess}(X)$$

X represents the original dataset, while X' is the preprocessed dataset. The preprocessing step ensures that the model can be trained on a high-quality dataset, thereby improving the accuracy of the recommendation. After data preprocessing, we use the embedding layer to map the sparse feature vector X to a dense embedding vector space. The role of the embedding layer is to convert high-dimensional sparse features into low-dimensional dense vectors to facilitate subsequent feature crossover operations. This step can be formally expressed as:

$$E(X') = W_{emb} \cdot X' + b_{emb}$$

Among them, $W_{emb}$ is a trainable weight matrix used to encode each feature into a vector of fixed length, and $b_{emb}$ is a bias term. The introduction of the embedding layer not only reduces the dimensionality difference between features, but also enhances the model's ability to capture complex relationships between features, so that subsequent feature cross-operations can more accurately reflect the user's real needs. After completing feature embedding, we introduced the Separate Embedding Cross Network, the core idea of which is to perform independent cross operations on features in different dimensions and explicitly control the depth of feature crossover by adjusting the number of network layers. The rationale for separating feature cross operations across different dimensions lies in its ability to explicitly control the complexity of interactions between features. Traditional methods often treat all features equally, applying a single cross operation across the entire feature set. In contrast, by isolating feature interactions in specific dimensions, our approach reduces noise and enhances the model's ability to identify meaningful patterns. This approach allows for more targeted interaction modeling, ensuring that features from different domains (e.g., user behavior, item characteristics) are combined in a way that better reflects their real-world interdependencies. This selective feature crossing results in more accurate predictions and improved recommendation quality. The main function of this model is to gradually increase the complexity of the relationship between features by stacking them layer by layer, so as to better simulate the diversity and complexity of user behaviors in the real world. This process can be expressed as:

$$H^{(i)} = f(W_C \cdot (H^{(i-1)} \otimes H^{(i-1)}) + W_R \cdot H^{(i-1)} + b_C)$$

Among them, $H^{(i)}$ represents the output matrix of layer i, and $H^{(i-1)}$ is the output of the previous layer. $W_C$ and $W_R$ are the weight matrices of the cross-layer and residual connection respectively, $b_C$ is the bias vector of the cross-layer, and $\otimes$ represents element-level multiplication. And $f(\cdot)$ is the activation function. In this way, we can not only effectively control the depth of feature crossover, but also reduce the computational complexity and improve the efficiency of the model. The design of the model enables the model to achieve faster training speed and lower memory usage while maintaining high accuracy. The overall architecture diagram is as Figure 1.

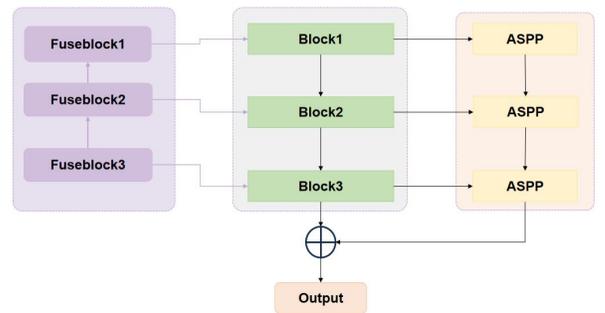

Figure 1 Overall network architecture

After the feature crossover is completed, we use sum pooling to aggregate feature information at different levels [15]. The purpose of this step is to simplify the model structure, reduce the number of parameters, and retain the information of important features [16]. The specific process of sum pooling can be expressed as:

$$P(H^{(L)}) = \sum_{i=1}^{N} h_i^{(L)}$$

Where N is the number of user features, L is the total number of layers in the network, and $h_i^{(L)}$ is the embedding vector of the i-th user in the L-th layer. The result of summing and pooling P will be used as the input of the prediction layer to further extract the importance of features and generate the final recommendation results. Through summing and pooling, we can effectively reduce the feature dimension while retaining the most representative information, which is crucial to improving the generalization ability and prediction performance of the model.

The prediction layer is the last link of the model, which maps the summed pooling results to the final predicted value. The prediction layer is usually a simple fully connected layer, which can be expressed as:

$$y = g(V \cdot P + b)$$

Where V and b are the weight and bias of the prediction layer, respectively, and $g(\cdot)$ can be a sigmoid function or other appropriate activation function, depending on the goal of the recommendation task. For example, in the click-through rate prediction scenario, $g(\cdot)$ outputs the probability that a user clicks on an ad. Through the above series of steps, our model can effectively learn the user's behavior patterns and make accurate recommendations based on them.

III. EXPERIMENT

A. Datasets

The datasets used in this paper are the Criteo dataset and the AutoML dataset. The following is a detailed introduction to the two datasets. The Criteo dataset is a public dataset widely used in click-through rate prediction research, which comes from an actual online advertising system. The dataset contains a large number of records of user click behaviors, including user characteristics, ad characteristics, and whether the user clicked on the ad. Specifically, the features in the Criteo dataset include but are not limited to the user's age, gender, geographic location, device information, historical behavior data, etc., as well as related attributes of the ad, such as ad category, ad size, display location, etc. These features are all pre-processed, such as anonymization to protect user privacy, and feature engineering to enhance the representation ability of the features. In the Criteo dataset, each sample corresponds to an ad display event, and the target variable is whether the user will click on the ad, which is usually expressed in binary form (click is 1, no click is 0).

The AutoML dataset originates from the field of automatic machine learning and is mainly used to evaluate automated machine learning processes, especially automated feature selection, model selection, and hyperparameter optimization. This dataset covers a variety of machine learning tasks, such as regression, classification, clustering, etc., and provides a rich combination of features and labels. One of the characteristics of the AutoML dataset is that it contains multiple sub-datasets, each of which reflects different types of tasks and application scenarios, such as image recognition, text classification, time series prediction, etc. What these sub-datasets have in common is that they are all carefully designed to evaluate the capabilities of automated machine learning frameworks.

B. Experimental Result

In order to verify the effectiveness of the model, the model is compared with the current mainstream deep learning recommendation model. The experimental results are as Table 1.

Table 1 Experiment result in Criteo

| Feature | Logloss | AUC |
|---|---|---|
| FM | 0.5233 | 0.7133 |
| NFM | 0.5123 | 0.7219 |
| AFM | 0.4972 | 0.7366 |
| DeepFM | 0.4922 | 0.7489 |
| XDeepFM | 0.4891 | 0.7541 |
| Ours | 0.4728 | 0.7623 |

Table 1 shows the performance results of different recommendation system models on the Criteol dataset, including two evaluation indicators: Logloss (logarithmic loss) and AUC (area under the curve). Logloss measures the difference between the model's predicted probability distribution and the actual label. The smaller its value, the more accurate the model's prediction; while AUC reflects the model's ability to distinguish between positive and negative samples. The closer the AUC value is to 1, the better the model's classification performance. As can be seen from the table, the Logloss and AUC values of all models are different, showing the difference in the prediction performance of different models.

First, the FM (Factorization Machine) model, as a classic factor decomposition machine model, has a Logloss value of 0.5233 and an AUC value of 0.7133, which shows that although the model has certain capabilities in feature crossover, its prediction performance is relatively limited when processing complex data sets. Subsequently, the NFM (Neural Factorization Machines) model improved the FM model by introducing a neural network structure, with the Logloss value reduced to 0.5123 and the AUC increased to 0.7219, showing the advantages of neural networks in feature interaction. The AFM (Attentional Factorization Machine) model further introduces the attention mechanism [17]. By assigning different weights to different feature combinations, its Logloss value is 0.4972 and its AUC value reaches 0.7366, which is a significant improvement over the previous model. The DeepFM model combines the advantages of deep learning and factorization machines, with a Logloss value of 0.4922 and an AUC value of 0.7489, showing the powerful ability of deep learning in feature representation learning. The XDeepFM model further improves the performance of the model by introducing an explicit feature cross-layer, with a Logloss value of 0.4891 and an AUC value of 0.7541, which shows that explicitly modeling the interaction between features can improve the prediction ability of the model. Finally, the model proposed in this paper (Ours) achieved the best results in both Logloss and AUC, with a Logloss value of 0.4728 and an AUC value of 0.7623, which shows that the design of the separation embedding cross network effectively improves the performance of the model in terms of feature cross and prediction accuracy.

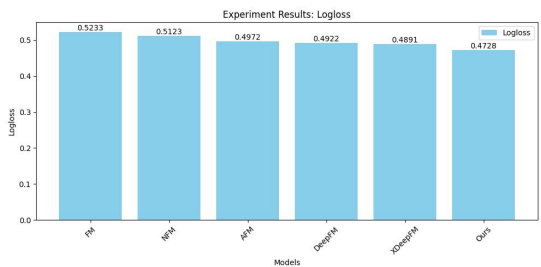

Figure 2 Result Logloss on Criteo

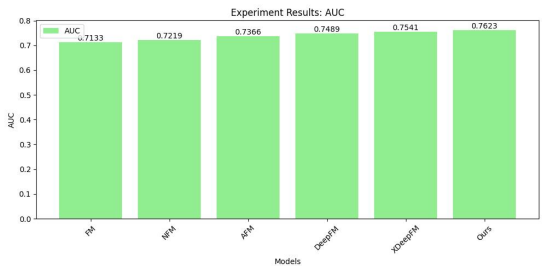

Figure 3 Result AUC on Criteo

As shown in Figures 2 and 3, we show the resulting bar graphs of AUC and logloss.

Table 2  Experiment result in AutoML

| Feature | Logloss | AUC |
| --- | --- | --- |
| FM | 0.1833 | 0.7321 |
| NFM | 0.1812 | 0.7433 |
| AFM | 0.1755 | 0.7514 |
| DeepFM | 0.1723 | 0.7621 |
| XDeepFM | 0.1703 | 0.7736 |
| Ours | 0.1691 | 0.7892 |

Table 2 shows the experimental results of different recommendation system models on the AutoML dataset, using Logloss (logarithmic loss) and AUC (area under the curve) to measure the performance of the model. From the perspective of Logloss, the value of the FM model is 0.1833, indicating that its prediction error is relatively large; the NFM model is slightly improved, with a Logloss value of 0.1812; the AFM model further reduces the Logloss value to 0.1755 by introducing the attention mechanism; the DeepFM model combines deep learning with factorization machines, and the Logloss value is reduced to 0.1723. The XDeepFM model further reduces the Logloss value to 0.1703 through an explicit feature cross-layer. The model proposed in this paper (Ours) shows the best Logloss value of only 0.1691, showing its advantage in prediction accuracy.

From the perspective of AUC, the FM model has an AUC value of 0.7321, showing general classification performance; the NFM model has an AUC value of 0.7433, which is slightly improved; the AFM model has an AUC value of 0.7514, showing the positive impact of the attention mechanism; the DeepFM model has an AUC value of 0.7621, which further improves the classification ability of the model; the XDeepFM model has an AUC value of 0.7736, indicating the effectiveness of explicit feature crossover; and the model proposed in this paper (Ours) has an AUC value of 0.7892, showing the strongest classification performance. Overall, the experimental results show that by introducing more advanced feature crossover and attention mechanisms, the prediction performance of the model on the AutoML dataset can be significantly improved. Similarly, in order to further demonstrate our results, a bar graph of the experimental results is given.

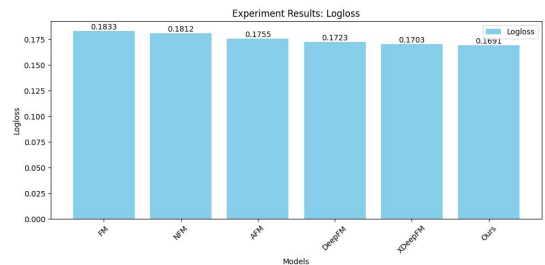

Figure 4 Result Logloss on AutoML

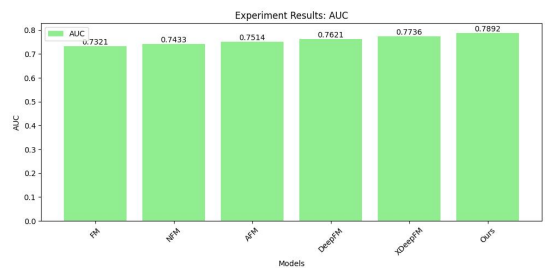

Figure 5 Result AUC on AutoML

As can be seen from Figures 4 and 5, our model also achieved the best results on AutoML.

## IV. CONCLUSION

Through the analysis of existing recommendation systems, we found that traditional methods have many limitations when facing big data and complex user behaviors. To this end, we developed an innovative recommendation system model that combines advanced feature cross-talk technology and attention mechanism to fully mine the embedded vector information and improve the accuracy of recommendations. Experimental results show that the proposed model can not only significantly improve the recommendation effect, but also effectively deal with the problems of high computational complexity and long feature selection time. In addition, the model shows good generalization ability and stability in practical applications, especially on large-scale datasets, it outperforms other mainstream models. Future research can further explore how to combine more diversified user behavior data in order to build a more personalized and intelligent recommendation system.